%% file: hopgnn-arxiv24.tex
\documentclass[sigplan,10pt]{acmart}
\renewcommand\footnotetextcopyrightpermission[1]{}

\usepackage{algorithm}
\usepackage{algorithmicx}
\usepackage{algpseudocode}
\algdef{SE}[DOWHILE]{Do}{doWhile}{\algorithmicdo}[1]{\algorithmicwhile\ #1}%
\usepackage{breqn}
\usepackage{makecell}
\usepackage[normalem]{ulem}
\usepackage{pifont}
\usepackage{subfigure}
\useunder{\uline}{\ul}{}

\usepackage{tikz}
\usepackage{amsmath}

\usepackage{filecontents}

\usepackage{graphicx}
\usepackage{multirow}
\usepackage{subfigure}

\usepackage{breakurl}

\usepackage{filecontents}
\usepackage{epstopdf}
\usepackage{rotating}
\usepackage{cleveref}
\usepackage{adjustbox}
\usepackage{comment}
\usepackage{pifont}

\crefformat{section}{\S#2#1#3} 
\crefformat{subsection}{\S#2#1#3}
\crefformat{subsubsection}{\S#2#1#3}

\crefname{section}{§}{§§}
\Crefname{section}{§}{§§}

\newcommand{\techa}{micrograph-based GNN training}
\newcommand{\techb}{pre-gathering}
\newcommand{\techc}{micrograph merging}
\newcommand{\techA}{Micrograph-Based GNN Training}

\graphicspath{{./fig/}}

\begin{document}

\newcommand{\pname}[1]{HopGNN{#1}}
\title{\pname{}: Boosting Distributed GNN Training Efficiency via Feature-Centric Model Migration}

\author{{Weijian Chen, Shuibing He, Haoyang Qu, and Xuechen Zhang $^{\dagger}$} \and {Zhejiang University, $^{\dagger}$ Washington State University Vancouver}}

\input{tex/abstract}


\settopmatter{printfolios=true}
\maketitle
\pagestyle{plain}

\input{tex/introduction}

\input{tex/background}
\input{tex/motivation}
\input{tex/micro}
\input{tex/design}

\input{tex/implementation}
\input{tex/eval}

\input{tex/discussion}

\input{tex/related}
\input{tex/conclusion}

\bibliographystyle{plain}
\bibliography{./bib/references}

\end{document}

%% file: tex/abstract.tex
\begin{abstract}
Distributed training of graph neural networks (GNNs) has become a crucial
technique for processing large graphs. 
Prevalent GNN frameworks are model-centric, necessitating the transfer of massive  
graph vertex features to GNN models, which leads to a significant
communication bottleneck. 
Recognizing that the model size is often significantly smaller than the feature size,
we propose \pname{}, a feature-centric framework that reverses this paradigm by
bringing GNN models to vertex features.
To make it truly effective,
we first propose a micrograph-based training strategy that trains the model
using a refined structure with superior locality to reduce remote feature retrieval.
Then, we devise  
a feature pre-gathering approach that merges multiple fetch operations into a
single one to eliminate redundant feature transmissions.
Finally, we employ a micrograph-based merging method that adjusts the number of micrographs
for each worker
to minimize kernel
switches and synchronization overhead. Our experimental results demonstrate that
\pname{} achieves a performance speedup of up to 4.2$\times$ compared to the
state-of-the-art method, namely $P^3$.
\end{abstract}

%% file: tex/introduction.tex
\section{Introduction}
\label{intro}

\noindent\textbf{Motivation.} 
The emerging graph neural networks (GNNs) are designed
for learning from graph-structured data.
They are widely employed in various graph-related tasks (e.g.,
vertex classification~\cite{hang2021collective, zhang2020every},
edge prediction~\cite{zhang2018link, zhang2023page}, and graph
classification~\cite{sui2022causal, bevilacqua2021size})
and have shown superior performance compared
to traditional graph algorithms in diverse domains, such as
recommendation systems~\cite{yang2021consisrec, wu2019session},
social networks analysis~\cite{zhang2022improving}, and
drug discovery~\cite{wang2022extending}.

Input graph datasets for GNN training consist of both
topology and vertex features~\cite{gnnlab-eurosys22, distdgl-ia320}.
The volume of real-world graph datasets can easily surpass the memory
capacity of a single machine. For example, the sizes of the
Pinterest~\cite{ying2018graph} and ByteDance~\cite{bgl-nsdi23}
datasets are 18 TB  and 100 TB respectively. Therefore, GNN models
are typically trained on distributed clusters. In distributed
GNN training, graph datasets are partitioned and distributed
across multiple servers. During each iteration, 
each worker on the server uses a subgraph as the input 
to train a local
copy of the GNN model. During the training, a large amount of
vertex features need to be fetched from remote servers, leading
to significant communication bottlenecks~\cite{p3-osdi21, bgl-nsdi23, distgnn-sc21}.

\noindent\textbf{Limitations of the state-of-the-art systems.} Many recent works
are proposed to reduce the time of remote feature fetching.
For the convenience of discussion, we name the existing approaches 
as \textit{model-centric} GNN frameworks in which vertex features are
moved to the GPU servers where GNN models are trained.
Specifically, ~\cite{pagraph-socc20, distgnn-sc21, bgl-nsdi23, distdgl-ia320} 
improve the hit rate of local features but compromise the
model accuracy (\cref{exp:acc}) using approximation-based methods. 
Such approaches are unsuitable for scenarios requiring high 
precision, as even 0.1\% accuracy drop in recommendation systems may lead 
to revenue losses of millions of dollars~\cite{persia-kdd22,
under-hpca21, checknrun-ndsi22,aibox-cikm19}.
Other studies~\cite{gnnlab-eurosys22, bgl-nsdi23, las-ics22} use GPU 
memory to cache popular vertex features. They are limited by the cache size
especially for large graphs.
To avoid remote feature fetching, $P^3$~\cite{p3-osdi21} combines 
model parallelism and data parallelism based on random hash 
partitioning. However, it is designed for GNNs 
that have small hidden dimensions and its performance gain is
reduced as the number of hidden layers increases~\cite{p3-osdi21, bgl-nsdi23}. 
Because of the deficiencies of the model-centric GNN frameworks, we 
need a novel solution, which provides high model accuracy and 
can be applied to a wide range of GNNs. 


\noindent\textbf{Our work.} In this paper, we propose \pname{}, \textit{the
first feature-centric} GNN framework that reverses the existing \textit{model-centric} paradigm by moving GNN models to the requested vertex features. 
\pname{} is motivated by the observation that the size of model parameters 
is significantly smaller than the volume of vertex features (\cref{motiv_bottleneck}), thus
transferring model parameters incurs less cost
than fetching graph features. 
However, a naive implementation of this framework could still result in
considerable data movement, due to the complex computation dependencies of GNN models 
and the need to transmit intermediate data, such as partial aggregation 
results and activations. In fact, despite being beneficial in certain
scenarios, it may increase the data movement by up to
2.59$\times$ compared to the model-centric approach (\cref{sec:challenges}).



To make \pname{} truly effective, 
we devise three optimization techniques.
First, we propose a new abstract \textit{micrograph} as the fundamental training
unit for each worker (\cref{design0}). 
As a more refined unit compared to the traditional \textit{subgraph}, the
micrograph offers superior data locality, i.e., there is a
higher
probability that the root vertex of a micrograph and its fanout neighbors reside
on the same GPU server. 
Consequently, micrograph-based training streamlines computational dependencies,
mitigates the production of intermediate data, and reduces remote feature retrieval (\cref{design1}).
Second, we devise a \textit{pre-gathering} method to elevate communication
efficiency (\cref{design2}). 
This approach consolidates multiple feature fetch operations into a single operation, thereby reducing redundant feature transmissions.
Finally, we introduce a \techc{} approach that adaptively allocates
micrographs to each worker (\cref{design3}) for reduced model migrations, thus
minimizing kernel switches and synchronization overhead.
%

\noindent\textbf{Contributions.} We make the following contributions:
\begin{itemize}
\item To the best of our knowledge, we are the first to propose
the feature-centric distributed GNN training strategy using model
migration to reduce the feature communication overhead.

\item We further analyze the new challenges introduced by the 
naive feature-centric approach and propose three techniques which 
collectively enhance its efficacy and practicality, without compromising model accuracy. 

\item We implement \pname{} on the widely used DGL~\cite{dgl-arxiv19} 
framework and conduct extensive experiments using five representative 
datasets with five GNN models on a distributed GPU cluster. 
The experimental results demonstrate 
that \pname{} can achieve 
up to 4.2$\times$ speedup compared to the the 
state-of-the-art counterpart, $P^3$~\cite{p3-osdi21}.
\end{itemize}

%% file: tex/background.tex
\section{Background}
\label{background}

This section uses a vertex classification task as an 
example to illustrate the basic concepts and training 
process of GNNs.



\begin{figure}
	\centering
	\includegraphics[width=3.4in]{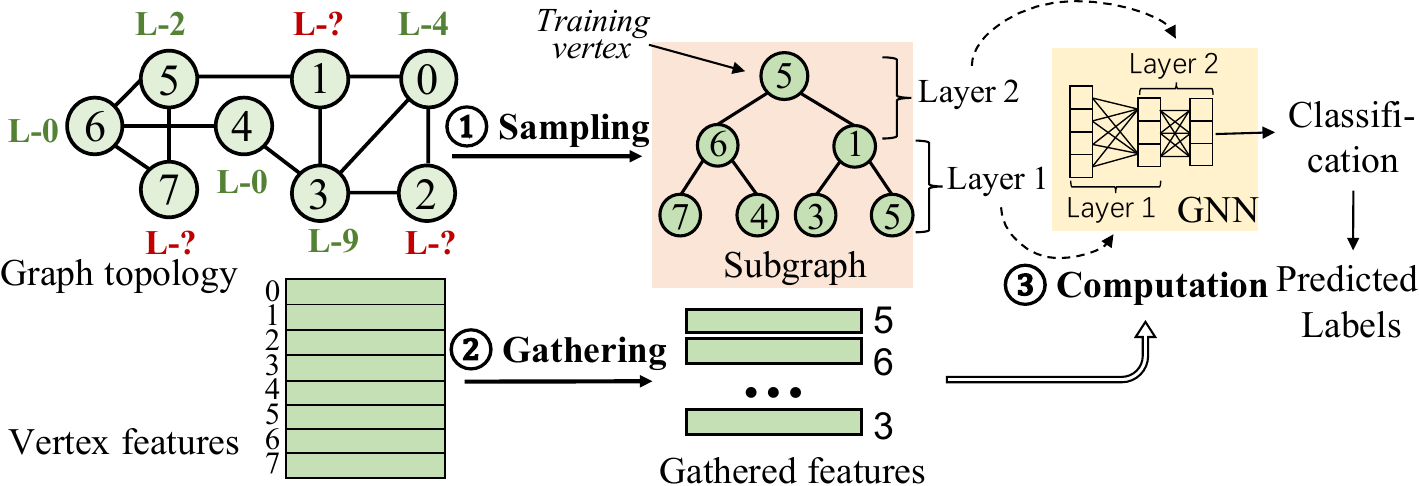}
	\caption{A GNN training example.}
	\label{fig:subgraph-gnn}
\end{figure}

\noindent\textbf{Input graph datasets.}
The input data for GNN training includes both the graph topology 
and the vertex features, as shown in Figure~\ref{fig:subgraph-gnn}.
In the example, we use a social network graph where each vertex
corresponds to a user and each edge represents a relationship between
two users. Each vertex is associated with a vertex feature, which is 
stored in an embedding vector. The embedding vectors can encode 
vertex features like age, gender, geographical
location, etc. The goal of the GNN task is to predict the preferred
topic for each user. Some users in the graph have revealed their
preferences on topics numbered from 0-9 (denoted as `L-x'), each of which represents a 
topic, e.g., sports, music, etc. We will use this disclosed information
as ground truth for training the GNN model. After training, 
the trained model is used to predict the preferred topics for users who 
have not disclosed their preferences.

\begin{figure}
	\centering
	\includegraphics[height=1.2in]{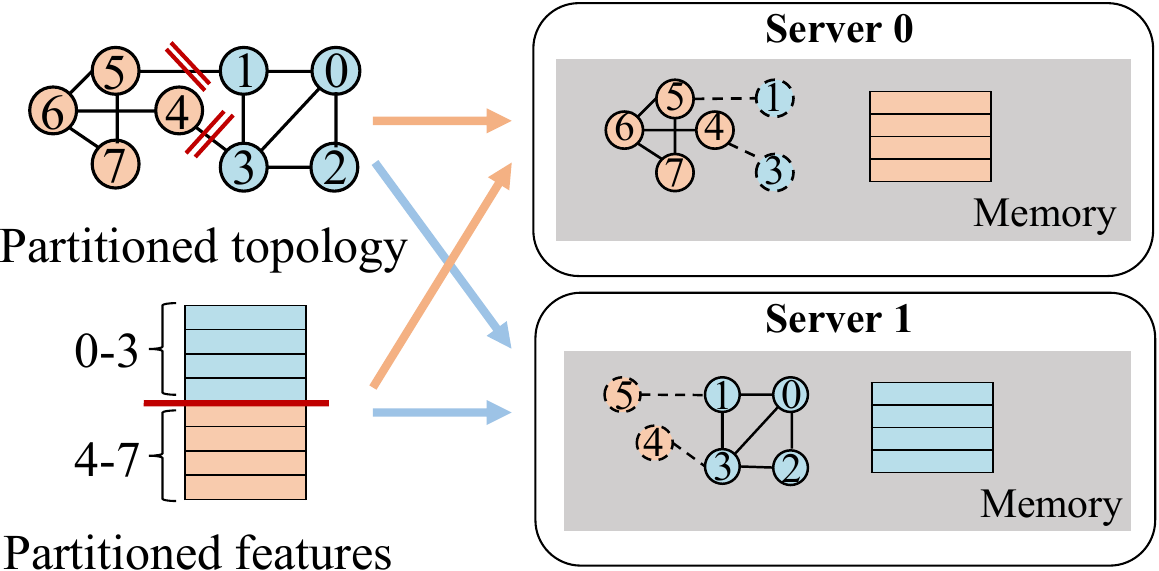}
	\caption{An example of partitioned graph topology and features on two GPU servers.}
	\label{fig:distgnn_partiton}
	\vspace{-0.1in}
\end{figure}


\begin{figure*}[!htb]
	\begin{minipage}[b]{0.6\textwidth}
		\flushleft
		\includegraphics[width=4.2in,height=2.1in]{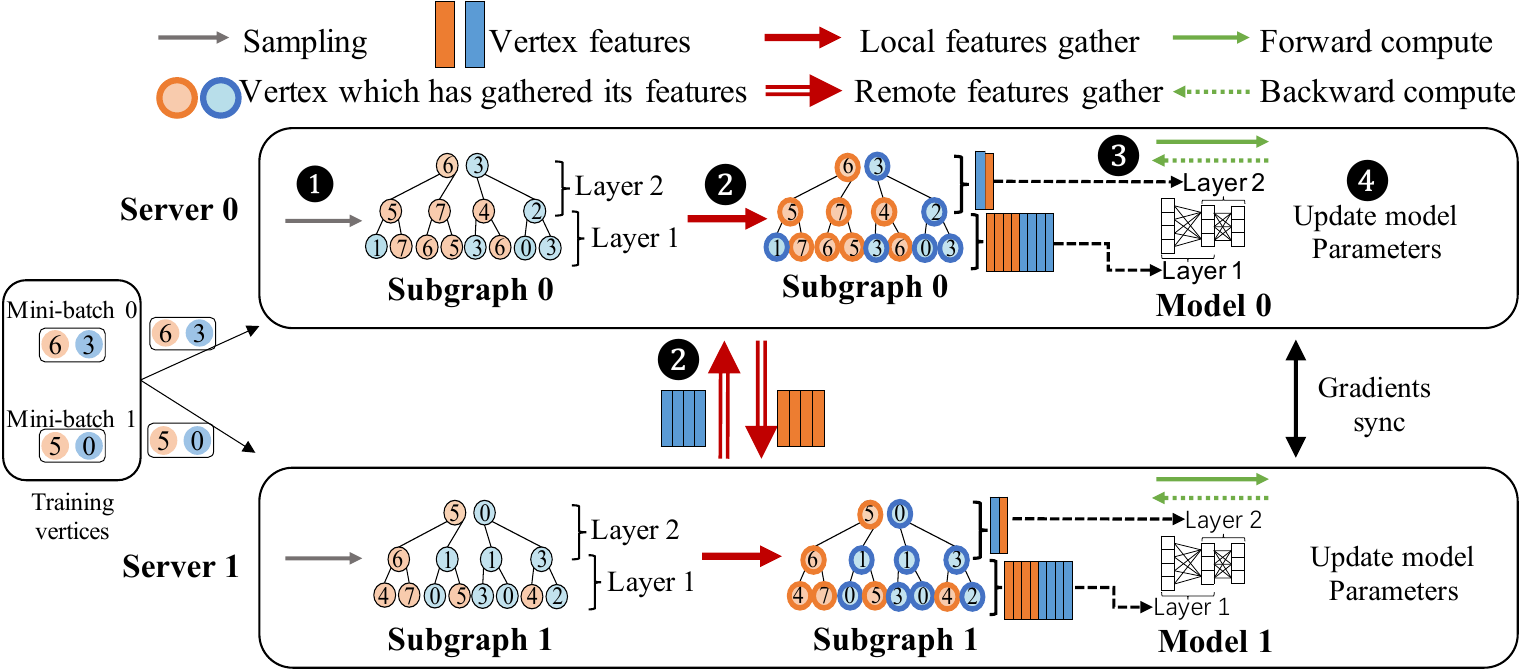}
		\caption{\footnotesize{An example of model-centric distributed GNN training approach.}}
		\label{fig:distgnn_default}
	\end{minipage}%
	\begin{minipage}[b]{0.4\textwidth}
		\flushright
		\includegraphics[width=2.6in, height=1.2in]{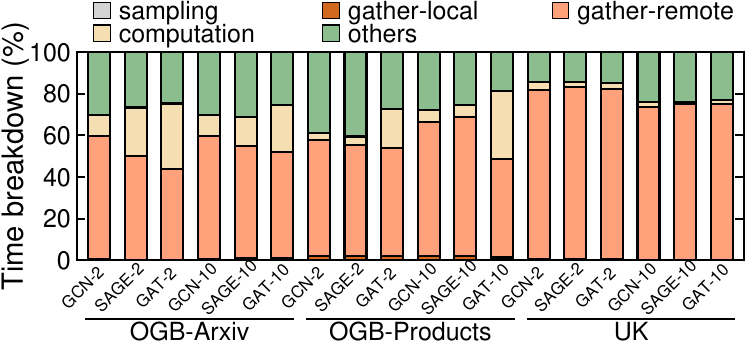}
		\vspace{-0.1in}
		\caption{\footnotesize{Training time breakdown. `Model-x' means the fanout is x.`SAGE' denotes GraphSAGE.}
		}
		\label{fig:bottleneck}
		\flushright
		\includegraphics[width=2.6in, height=0.6in]{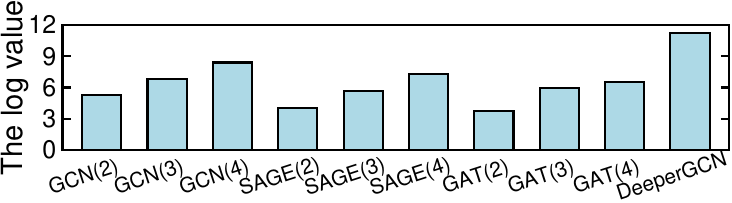}
		\vspace{-0.15in}
		\caption{\footnotesize{The $\alpha$ value of different models. `Model (x)' denotes that the number of model layers is x.}}
		\label{fig:motiv_ratio}
	\end{minipage}%
\end{figure*}


\noindent\textbf{Subgraph-based GNN training on a single GPU server.}
GNNs leverage labeled vertices, known as training vertices, to train a
multi-layer neural network model across multiple epochs. 
Similar to traditional DNN
training, each epoch involves multiple iterations to process all training
vertices once. 
Each iteration randomly selects a batch of training vertices and consists of
three key steps. 


The \textit{sampling} step involves k-hop neighbor sampling from the training vertices to
generate a k-layer \textit{subgraph}, where k equals the number of model layers. The
user-defined parameter `fanout' dictates the number of neighbors sampled per
vertex. For instance, the subgraph depicted in Figure~\ref{fig:subgraph-gnn} is
produced by a 2-hop neighborhood sample from a single training vertex 5 with a
fanout of 2. A subgraph may encompass a batch of training vertices, as shown in
Figure~\ref{fig:distgnn_default}.
The \textit{gathering} collects the features of each vertex within the constructed subgraph.
The \textit{computation} processes the subgraph layer by layer, beginning with the first layer. For each layer, aggregation operations (like addition or averaging) are conducted on the neighboring features of each vertex. Subsequently, a neural network transformation updates the feature representations. The updated features of the training vertices are fed into a classification network layer to produce predicted labels. Using the true labels, a backward propagation step then follows to refine the model parameters.

\noindent\textbf{Distributed GNN training.}
For large-scale graphs that exceed the storage capacity of a single GPU
server, it is necessary to distribute the graph's topology and vertex features
across multiple servers. 
Figure~\ref{fig:distgnn_partiton} depicts  
the process of partitioning the graph into two parts (denoted by two colors) for
distributed GNN training on 
two servers. 
Given that the size of the graph topology is smaller than
that of the vertex feature embeddings 
(for instance, 6 GB for topology vs. 53 GB for vertex features in the OGB-Papers100M dataset), 
several studies~\cite{pagraph-socc20, distdgl-ia320, aligraph-kdd19} opt to redundantly store a subset or 
the entire topology (e.g. vertices 1 and 3 on server 0) in a small portion of
the host memory.
This strategy aims to minimize data transmission during the sampling process. 

Figure~\ref{fig:distgnn_default} shows a typical distributed GNN training
approach that utilizes data parallelism. Each
server hosts a complete GNN model copy.
At the beginning of each iteration, each model 
is randomly assigned a disjoint mini-batch of vertices for training. 
For example, Model $0$ is allocated the training 
vertices $\{6, 3\}$, and model $1$ receives $\{5, 0\}$. Subsequently,
the training processes independently perform the 2-hop
sampling(\ding{182}), feature gathering from local or remote servers
(\ding{183}), and 
computation steps(\ding{184}). Finally, the parameter gradients from different 
models are synchronized, and the model parameters are updated (\ding{185}). Since the models remain stationary on their respective servers without migration throughout the training process, this method is characterized as \textit{model-centric}.


%

%% file: tex/motivation.tex
\section{Motivation and Challenges}
\label{motiv}

\subsection{Communication Bottleneck 
in GNN Training}
\label{motiv_bottleneck}

\begin{figure*}[!htb]
	\begin{minipage}[b]{0.65\textwidth}
		\centering
		\includegraphics[width=4.5in, height=2in]{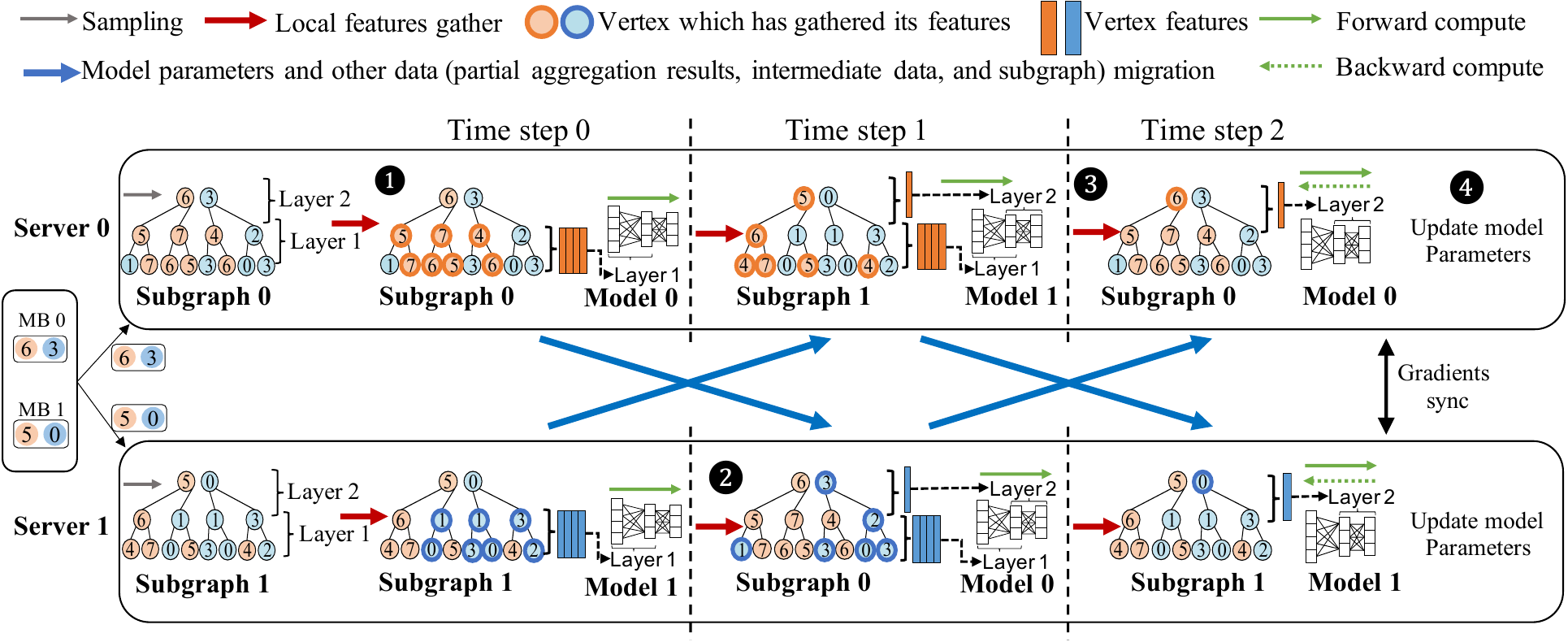}
		\caption{The naive feature-centric approach.}
		\label{fig:naive_migrate_v0}
	\end{minipage}%
	\begin{minipage}[b]{0.35\textwidth}
		\centering
		\includegraphics[width=2.4in, height=1.2in]{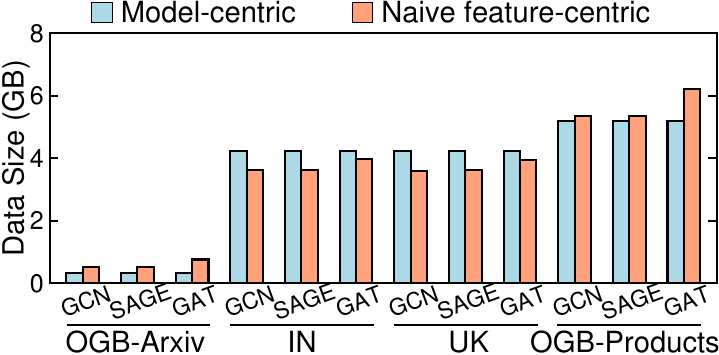}
		\caption{Comparison of the amount of transferred data: 
			model-centric training vs. naive feature-centric training.
			The y-values are log2 scaled.}
		\label{fig:naive_trans}
	\end{minipage}%
\end{figure*}

In this section, we study the performance of GNN training using
DGL~\cite{dgl-arxiv19} which is a widely used GNN framework in industry. 
We train three popular GNN models (GCN~\cite{gcn-iclr17},
GraphSAGE~\cite{sage-nips17}, and GAT~\cite{gat-iclr18}) on three graph
datasets~\cite{ogb-arxiv20} (OGB-Arxiv, OGB-Products, and UK). 
The fanout is two or ten and the number of GNN layers is
three, following the settings in the previous works~\cite{pagraph-socc20,
gnnlab-eurosys22}. 
The detailed evaluation methodology is  described in \cref{eval}. 

\noindent\textbf{Observation I:
Vertex feature gathering causes the communication bottleneck.} 
We utilize PyTorch Profiler~\cite{profiler} to collect time metrics and present
the detailed breakdown of the execution times in
Figure~\ref{fig:bottleneck}. Notably, gathering remote vertex features consumes between 44\% to 83\% of the
total training time. In comparison, the combined
time for sampling and computation stands at an average of merely 11\%. Despite
the framework's ability to parallelize graph sampling and computation via GPUs,
it falls short in mitigating the time required for inter-server communication,
particularly when transferring substantial volumes of vertex features. For
instance, with GAT~\cite{gat-iclr18} on the OGB-Products dataset, a significant
35 GB of vertex features are exchanged per epoch, contrasting sharply with the
0.4 GB of graph topology data. This analysis underscores that remote feature
gathering is the predominant factor influencing the end-to-end training time in
distributed GNN training.

\noindent\textbf{Observation II: The volume of data transferred for vertex feature gathering is substantially greater than the size of the model.}
To quantify this, we introduce a ratio 
$\alpha$, representing the amount of training data fetched from remote servers
per iteration relative to the size of the model parameters. 
This ratio was measured across prevalent GNN models with various number of layers.
Figure~\ref{fig:motiv_ratio} presents the findings,
with the y-axis depicting $log_2\alpha$ to accommodate the vast range of
values.
We observe that $\alpha$ varies
from 13.4 to 2368.1. Notably, in sophisticated deep GNN architectures, such as
DeeperGCN~\cite{deepergcn-arxiv20} with 112 layers, $\alpha$ reaches an
extraordinary 2368.1. This disparity stems from the fact that the number of
vertices within a subgraph increases more rapidly than the number of model
parameters, a consequence of k-hop sampling where each model layer corresponds
to a layer of the subgraph.

Based on these observations, we are inspired to leverage model migration to reduce the amount of data transfers during GNN training. Our goal is to move the model to GPU 
servers where the vertex features are located, rather than 
fetching features from remote servers.
For the sake of clarity, we term this the ``\textit{feature-centric}'' approach.

\subsection{A Naive Feature-Centric Training Approach}
\label{sec:challenges}

\noindent\textbf{A naive feature-centric approach.} It involves migrating the model to remote GPU
servers when the vertex features needed for a subgraph are not locally
available. However, this method necessitates transferring considerable amount of
intermediate data along with the model, owing to the computational dependencies
intrinsic to GNNs. Specifically, aggregation operations must be finalized after
acquiring the vertex features of all fanout neighbors. Additionally, backward
propagation is contingent on intermediate data produced during the forward pass,
and the computation of subgraphs must proceed sequentially, layer by layer.

Figure~\ref{fig:naive_migrate_v0} shows an example of this method when training
two mini-batches in Figure~\ref{fig:distgnn_default}.
We focus on a single iteration on model 0 for brevity. 
The process comprises three time steps.
At time step 0, model 0 initiates computation on layer 1 of subgraph 0. It gathers the features of vertices 4, 5, 6, and 7 locally and then inputs these into the model's first layer for forward propagation (\ding{182}). However, the features for vertices 0, 1, 2, and 3 in layer 1 of subgraph 0 are not locally available, resulting in a partial completion of layer 1's forward computation. We save the partial aggregation results and intermediate data within model 0 for temporary storage.
At time step 1, model 0, with the temporarily stored data and the topology of subgraph 0, migrates to server 1. There, it gathers the features of vertices 0, 1, 2, and 3. Consequently, the forward computation for layer 1 is fully executed, and layer 2's forward computation is partially completed (\ding{183}).
At time step 2, model 0 returns from server 1 to server 0, carrying the stored
data and the topology of subgraph 0. It gathers the features of the root vertex,
6, and uses the previously stored intermediate data to complete the forward and
backward computations for the entire GNN model (\ding{184}). Subsequently, model
0 synchronizes gradients with model 1 and performs parameter updates
(\ding{185}).

\noindent\textbf{Challenges.} While the naive feature-centric approach eliminates the need for remote
feature gathering, it may compromise performance due to extensive intermediate
data communication and frequent model transfers. Figure~\ref{fig:naive_trans}
compares the total data transmissions of the model-centric and naive
feature-centric methods. It shows that, despite being beneficial in certain
scenarios, the naive feature-centric approach can demand up to 2.59$\times$ the data communication
of the model-centric one. This significant communication overhead often results
in suboptimal performance for the naive feature-centric strategy.

To fully capitalize on the unique characteristic of GNNs, where model sizes are typically smaller than those of the remote vertex features, a more sophisticated approach is essential. This approach should facilitate model migration while mitigating the high communication overhead stemming from the computational dependencies inherent in GNNs.

%% file: tex/micro.tex
\section{New Abstraction: Micrograph}
\label{design0}

A primary source of inefficiency in the naive feature-centric approach lies in
its fundamental training unit: the subgraph. There is weak data
locality when retrieving features for the subgraph within a distributed
environment. To address this, we introduce the concept of a \textbf{micrograph},
a more refined data structure. With its enhanced data locality, as detailed
subsequently, the micrograph significantly diminishes the need for extensive
intermediate data communication and frequent model transfers.

\begin{figure}
	\centering
	\subfigure[]{
              \includegraphics[width=1.5in]{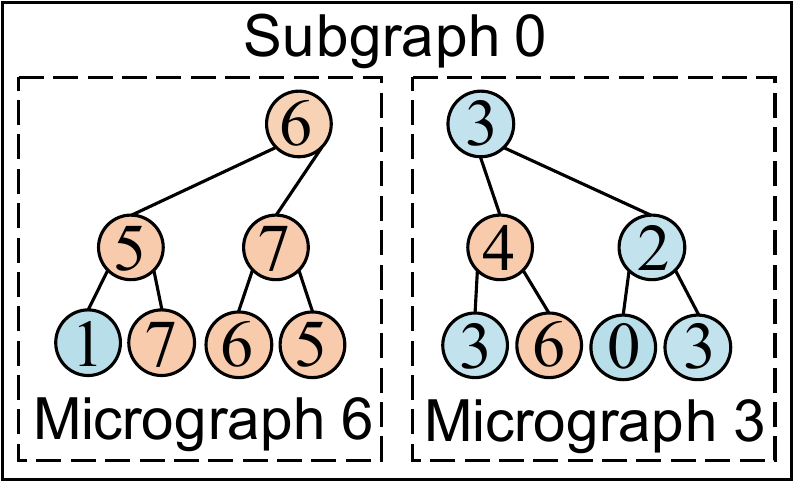}
              \label{fig:sub_micro_a}
      }
      \hspace{0.06in}
      \subfigure[]{
      	\includegraphics[width=1.5in]{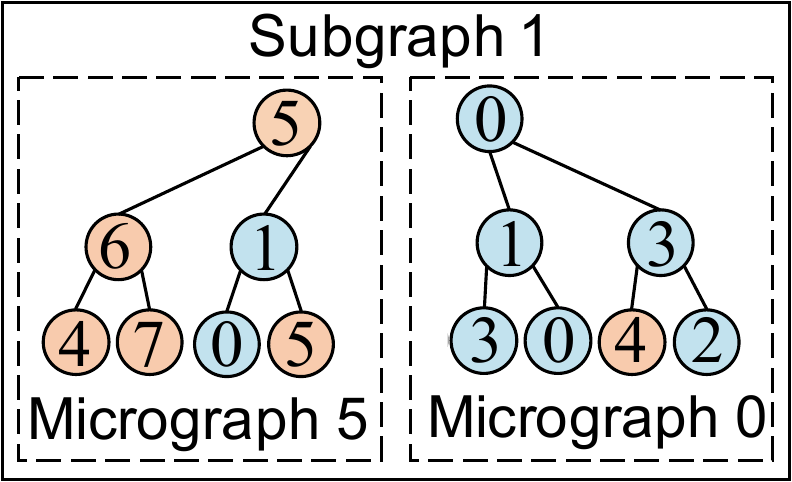}
      	\label{fig:sub_micro_b}
      }
      \vspace{-0.2in}
      \caption{Examples of subgraphs and micrographs.}
	\label{fig:sub_micro}
\end{figure}

\noindent\textbf{Micrograph definition.}
For each vertex $v$ in a mini-batch,
a micrograph is constructed using k-hop sampling to form a computation graph.
For convenience, we use the vertex ID to represent its corresponding
micrograph. A subgraph can encompass multiple
micrographs, as illustrated in Figure~\ref{fig:sub_micro}.
We assume that (1) the mini-batch size is two, (2) vertex features and graph topology data
are distributed across two GPU servers, and (3)
the fanout and the number of computation layers are both set to two.
Subgraph 0,
associated with mini-batch \{6, 3\}, comprises micrographs 6 and 3. Similarly,
subgraph 1, corresponding to mini-batch \{5, 0\}, includes micrographs 5 and 0.




\noindent\textbf{Data locality in micrographs.}
Micrographs, due to their finer granularity compared to subgraphs, inherently
exhibit superior data locality. Specifically, when employing widely-used graph
partitioning algorithms~\cite{metis98, dgcl-eurosys21, neugraph-atc19,pagraph-socc20, bgl-nsdi23, bytegnn-vldb22}, there is a high probability that the root
vertex of a micrograph and its fanout neighbors reside within the same
partition. Consequently, the server hosting the root vertex's features is also
likely to hold the features for its neighboring vertices

We use Figure~\ref{fig:sub_micro} as an example for illustration. 
For clarification, distinct colors are used to indicate the home locations of vertices; for
instance, red represents server 0, and blue represents server 1. In
Figure~\ref{fig:sub_micro_a}, we
need to retrieve the subgraph for the mini-batch \{6, 3\}. Consequently, vertices 6 and 3 are designated as the roots of micrographs 6
and 3, respectively. Leveraging data locality, micrograph 6 retrieves vertices
6, 5, and 7 from server 0. This process accesses 75\% of the feature vectors
needed for training micrograph 6, as three out of four vectors are read from
server 0. Similarly, micrograph 3, when trained, accesses 60\% of its required
feature vectors, with three out of five vectors read from server 1. It's
noteworthy that this feature locality is also present in micrograph 5 and 0.


\begin{table}[]
	\small
	\scalebox{0.9}{
		\begin{tabular}{|c|c|cccc|cccc|c|}
			\hline
			\multirow{3}{*}{\textbf{\begin{tabular}[c]{@{}c@{}}Sam-\\ pling\end{tabular}}}  & \multirow{3}{*}{\textbf{\#S}} & \multicolumn{4}{c|}{\textbf{METIS (\%)}}                                                                               & \multicolumn{4}{c|}{\textbf{Heuristic (\%)}}                                                                           & \multirow{3}{*}{\textbf{\begin{tabular}[c]{@{}c@{}}$R_{sub}$(\%)\end{tabular}}} \\ \cline{3-10}
			&                               & \multicolumn{2}{c|}{\textbf{Arxiv}}                                  & \multicolumn{2}{c|}{\textbf{Products}}          & \multicolumn{2}{c|}{\textbf{Papers}}                                 & \multicolumn{2}{c|}{\textbf{IT}}                &                                                                              \\ \cline{3-10}
			&                               & \multicolumn{1}{c|}{\textbf{2L}} & \multicolumn{1}{c|}{\textbf{10L}} & \multicolumn{1}{c|}{\textbf{2L}} & \textbf{10L} & \multicolumn{1}{c|}{\textbf{2L}} & \multicolumn{1}{c|}{\textbf{10L}} & \multicolumn{1}{c|}{\textbf{2L}} & \textbf{10L} &                                                                              \\ \hline
			\multirow{4}{*}{\textbf{\begin{tabular}[c]{@{}c@{}}Node-\\ wise\end{tabular}}}  & \textbf{2}                    & \multicolumn{1}{c|}{75}          & \multicolumn{1}{c|}{73}           & \multicolumn{1}{c|}{95}          & 88           & \multicolumn{1}{c|}{93}          & \multicolumn{1}{c|}{61}           & \multicolumn{1}{c|}{66}          & 64           & 50                                                                           \\ \cline{2-11} 
			& \textbf{4}                    & \multicolumn{1}{c|}{66}          & \multicolumn{1}{c|}{45}           & \multicolumn{1}{c|}{92}          & 79           & \multicolumn{1}{c|}{89}          & \multicolumn{1}{c|}{43}           & \multicolumn{1}{c|}{54}          & 46           & 25                                                                           \\ \cline{2-11} 
			& \textbf{8}                    & \multicolumn{1}{c|}{59}          & \multicolumn{1}{c|}{27}           & \multicolumn{1}{c|}{88}          & 68          & \multicolumn{1}{c|}{84}          & \multicolumn{1}{c|}{35}           & \multicolumn{1}{c|}{48}         & 36           & 12                                                                         \\ \cline{2-11} 
			& \textbf{16}                   & \multicolumn{1}{c|}{63}          & \multicolumn{1}{c|}{35}           & \multicolumn{1}{c|}{86}          & 61           & \multicolumn{1}{c|}{84}          & \multicolumn{1}{c|}{30}           & \multicolumn{1}{c|}{46}          & 32           & 6                                                                         \\ \hline
			\multirow{4}{*}{\textbf{\begin{tabular}[c]{@{}c@{}}Layer-\\ wise\end{tabular}}} & \textbf{2}                    & \multicolumn{1}{c|}{79}          & \multicolumn{1}{c|}{54}           & \multicolumn{1}{c|}{55}          & 52           & \multicolumn{1}{c|}{85}          & \multicolumn{1}{c|}{58}           & \multicolumn{1}{c|}{80}          & 53           & 50                                                                           \\ \cline{2-11} 
			& \textbf{4}                    & \multicolumn{1}{c|}{70}          & \multicolumn{1}{c|}{30}           & \multicolumn{1}{c|}{34}          & 28           & \multicolumn{1}{c|}{77}          & \multicolumn{1}{c|}{31}           & \multicolumn{1}{c|}{67}          & 30           & 25                                                                           \\ \cline{2-11} 
			& \textbf{8}                    & \multicolumn{1}{c|}{65}          & \multicolumn{1}{c|}{18}           & \multicolumn{1}{c|}{25}          & 14           & \multicolumn{1}{c|}{56}          & \multicolumn{1}{c|}{24}           & \multicolumn{1}{c|}{63}          & 18           & 12                                                                         \\ \cline{2-11} 
			& \textbf{16}                   & \multicolumn{1}{c|}{61}          & \multicolumn{1}{c|}{12}           & \multicolumn{1}{c|}{21}          & 9            & \multicolumn{1}{c|}{57}          & \multicolumn{1}{c|}{12}           & \multicolumn{1}{c|}{61}          & 12           & 6                                                                         \\ \hline
		\end{tabular}
	}
	\caption{Data locality of micrographs with various sampling and graph partition algorithms and model layers.}
	\label{table:same_ratio}
	\vspace{-0.25in}
\end{table}


To demonstrate the generality of this observation, we conduct 
experiments on four real-world open-source graph datasets. 
The first two datasets are partitioned with METIS~\cite{metis98} 
used in DGL~\cite{dgl-arxiv19}, and the last two large datasets are partitioned with 
a heuristic algorithm, as utilized in BGL~\cite{bgl-nsdi23} because 
the METIS algorithm runs out of memory when partitioning these two
graphs.
We utilize both the node-wise~\cite{sage-nips17} and layer-wise
~\cite{fastgcn-iclr18} random sampling algorithms.
We vary the number of servers (\#S) from 2 to 16 and 
observe the locality of micrographs for 
both shallow-layer GNNs (i.e., two layers, denoted as '2L') and 
deep-layer GNNs (i.e., ten layers, denoted as '10L').

\begin{figure*}
        \centering
        \includegraphics[width=7in, height=2.5in]{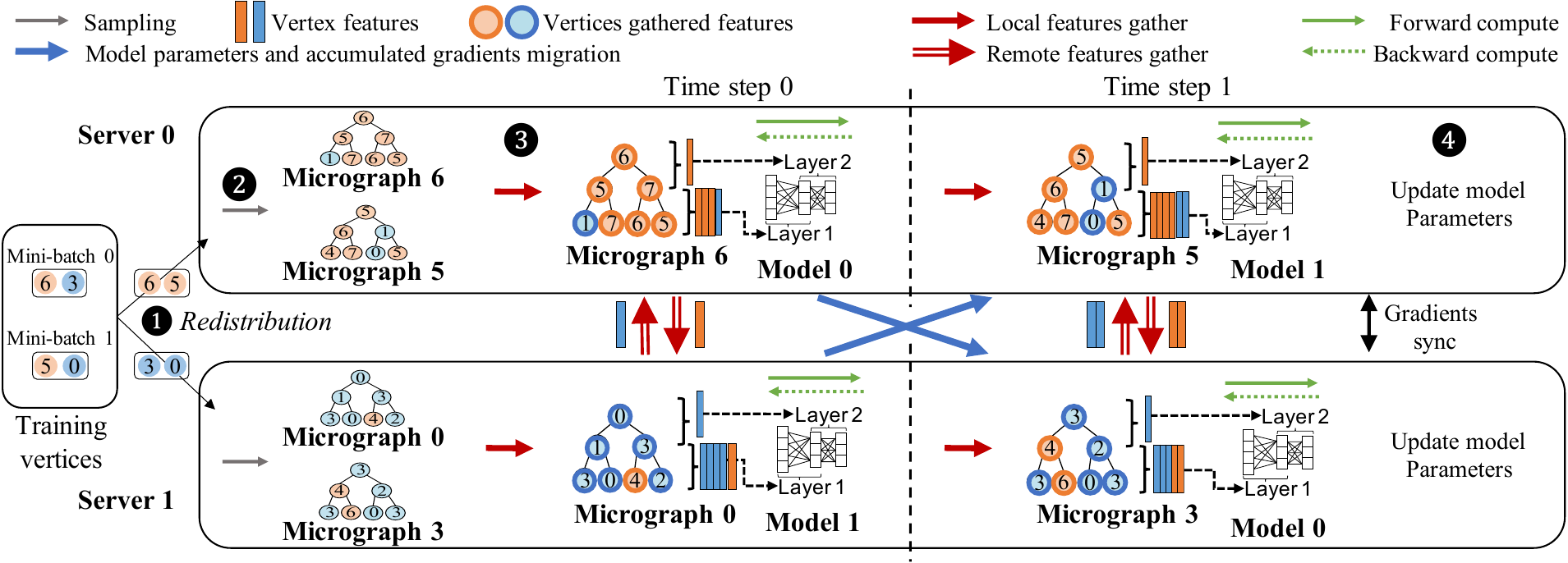}
        \caption{Example of training two mini-batches in one iteration in \pname{}.}
        \label{fig:keyidea}
\end{figure*}

We collect the number ($N_{colocated}$) of non-root vertices 
which are co-located with its root vertex in a micrograph 
during GNN training. Then, we compute the ratio ($R_{micro}$) 
of $N_{colocated}$ and $N_{total}$, where $N_{total}$ is the 
total number of vertices in a micrograph. We show $R_{mico}$ 
in Table~\ref{table:same_ratio}.
Likewise, we calculate the locality ($R_{sub}$) of subgraphs by 
dividing the count of non-root vertices co-located with a specified root vertex 
by the total number of vertices in the subgraph. 
For ease of presentation, we only show the mean value of $R_{sub}$.
We observe that $R_{micro}$ is consistently larger than 
$R_{sub}$, meaning micrographs' better locality. 
Furthermore, as the number of GPU servers 
increases from 2 to 16, the difference between $R_{micro}$ 
and $R_{sub}$ is increased from 1.59$\times$ to 10.60$\times$.


We credit this enhanced data locality to the prevalent graph partitioning
strategies employed in GNNs, including algorithms like METIS~\cite{metis98,
dgcl-eurosys21, neugraph-atc19} and GNN-specific
heuristics~\cite{pagraph-socc20, bgl-nsdi23, bytegnn-vldb22}. These methods
prioritize minimizing cross-machine feature transmission by strategically
assigning neighboring vertices to the same partition, thus promoting data
locality. As a result, when performing k-hop neighbor sampling from a root
vertex, the majority of vertices within the same micrograph are likely to be
co-located with the root on the same GPU server.

%% file: tex/design.tex
\section{Design of \pname{}}
\label{design_mainpart}
In this section, we detail the key ideas of \pname{} that tackles the
aforementioned challenges by leveraging data locality in micrographs. 
Subsequently, we present two enhancements designed to augment the
performance of \pname{}.


\input{tex/design1}
\input{tex/design2}
\input{tex/design3}

%% file: tex/design1.tex
\subsection{\techA{}}
\label{design1}

\textbf{Key idea.} 
We propose a novel approach, \techa{}, which decomposes a subgraph into multiple
micrographs and executes the complete forward and backward computations for each
micrograph on a single GPU server. During training, vertex features not
available locally are fetched from remote servers. This micrograph-based GNN
training offers two significant advantages: (1) It minimizes remote feature
gathering by exploiting data locality within micrographs. (2)
It eliminates intermediate data retrieval. Since micrographs are processed on a
single server, both forward and backward propagation can be completed once
vertex features are gathered in each layer of the micrographs.

\noindent\textbf{The procedure of micrograph-based GNN training.}
Assume that there are $N$ GPU servers.
Before training starts at each iteration, 
each server $s$ is assigned a model $d$
where $s \in [0, N-1]$ and $d$ is equal to $s$. 
Each model is randomly assigned a mini-batch of training 
vertices. 
We assign a home server for each vertex in the mini-batch
based on where the features for the vertex is located.
When training begins, \techa{} consists of the following four steps.
\textbf{(1) Redistribution of root vertices}. We will group 
root vertices in all mini-batches base on their home server IDs for task redistribution.
Then, each vertex group is assigned to the worker running
on the corresponding home server.
Since root vertices are randomly sampled from the global graph, 
the number of vertices received by each server is approximately equal at most cases. 
For instance, when we use four GPU servers for training, 
the load difference among them is less than 10\% for 97.3\%
training iterations for the four datasets used in our experiments.
\textbf{(2) Generating micrographs}. After the redistribution,
each server $i$ needs to use k-hop sampling to generate
a single set ${mg_{d}^{i}}$ of micrographs for each vertex group, 
where $d$ is determined by looking up the GNN model which
is originally selected to be trained using the vertex.
\textbf{(3) Training in $N$ time steps}. At time step 
$t$, model-$d$ migrates to server $s_{new} = (d+t) \% N$ 
and trains using micrographs $mg_{d}^{s_{new}}$. If the vertex 
features of the current micrograph are not available 
on the local server, they  will be fetched from the remote ones. 
To ensure model parameters are updated only after the whole 
subgraph training is completed, the temporary gradients 
obtained by training one micrograph is accumulated. 
\textbf{(4) Updating model parameters.} When the training on 
the last micrograph of the subgraph 
is completed, the accumulated gradients are synchronized 
among all GPU servers. 
Finally, the model parameters are 
updated to finish training for one iteration.


Figure~\ref{fig:keyidea} shows an example of our \techa{} for  two mini-batches
in one iteration on two GPU servers. 
Initially, both server 0 and 1 duplicate the DNN model. Eight features are
evenly distributed between these servers.
Then, training vertices are reassigned, with server 0 obtaining vertices 6 and
5, and server 1 getting vertices 0 and 3 (\ding{182}).
Next, each server independently generates micrographs through sampling (\ding{183}).
The training process is then divided
into two time steps (\ding{184}). During the first step, servers 0 and 1 train using micrographs 6 and 0, respectively. Upon completing the backward computation, intermediate data is discarded, retaining only the accumulated gradients.
Models are then migrated with their gradients: model 0 moves to server 1, and
model 1 to server 0.
In the second time step, the servers continue training using micrographs 5 and 3, respectively.
Finally, the gradients from both models are averaged, and parameter updates are
conducted (\ding{185}). This micrograph-based approach, as opposed to the 
subgraph-based training, reduces the transmission of vertex features between
servers through strategic model migration (8 features in
Figure~\ref{fig:distgnn_default} vs. 6 in Figure~\ref{fig:keyidea}).


\noindent\textbf{Limitations of the locality-optimized approach.} One might think that training models on redistributed micrographs without model
migration—such as model 0 on micrographs 6 and 5, and model 1 on micrographs 0
and 3—could enhance feature locality. 
While it is true, 
this approach could inadvertently
disrupt the training sequence for each model, as the sequence would be
randomized only within a local context rather than globally. For instance, model
0 would never be exposed to micrograph 3, as its features reside on a separate
server 1. This locality-optimized approach could introduce bias into the
mini-batch training data, potentially degrading the model's accuracy, as
discussed in \cite{globally-ipdps22,accele-hipc19}. Conversely, the
micrograph-based training method preserves model accuracy by maintaining the
globally randomized data sequence. For instance, model 0 consistently trains on
micrographs 6 and 3, matching the composition of the original mini-batch 0.
Additionally, the use of gradient accumulation, as shown in prior
research~\cite{zhang2023adam,gpipe-nips19,betty-asplos23}, does not compromise
training accuracy. The impacts of the locality-optimized approach on model 
accuracy are discussed in~\cref{exp:acc}.

A special case arises when the number of micrographs obtained from a subgraph is
less than the number of GPU servers in a cluster. This implies that on certain
machines there are no corresponding micrographs to train. In such cases, we
allow the model to do nothing on those machines until other models have
completed the corresponding micrographs' training.

%% file: tex/design2.tex
\subsection{Vertex Feature Pre-Gathering}
\label{design2}
Although \techa{} significantly diminishes the need for remote feature retrieval
by leveraging data locality within micrographs, it can inadvertently lead to
redundant transmissions of vertex features across consecutive time steps,
potentially resulting in suboptimal performance. We use the example in
Figure~\ref{fig:keyidea} to illustrate this.
Utilizing \techa{}, the worker on server 0 at time step 0 must
obtain the feature of vertex 1 from server 1. Upon completing the
computations for that time step, the vertex feature memory is cleared to prevent
GPU memory overflow. However, at time step 1, the worker on server 0 is required
to retrieve the features for both vertices 1 and 0 from server 1.
Consequently, for processing just two micrographs, server 0 ends up fetching a
total of three features from server 1, including a redundant transmission
for the feature of vertex 1.

\noindent\textbf{Key idea.}
We introduce vertex feature pre-gathering to mitigate redundant transmissions.
This approach capitalizes on the predictability of which vertices from the
micrographs will undergo training on a given server, regardless of the specific
models involved (e.g., model 0 or 1). For instance, referring to
Figure~\ref{fig:keyidea}, we can anticipate that at time step 0, vertex 1 will
be utilized by micrograph 6, and at time step 1, both vertices 1 and 0 will be
utilized by micrograph 5. Pre-gathering allows us to fetch the features
for vertex 1 and 0 from server 1 to server 0 in a single batch, thereby reducing
the communication cost from three feature transmissions without
pre-gathering to just two.

\noindent\textbf{Space overhead.} While pre-gathering additional non-local features could further minimize redundant transmissions, it necessitates extra memory space. To manage this, we limit pre-gathering to the features of vertices required for a single iteration of GNN training. As detailed in~\cref{design0}, due to the feature locality inherent in micrographs, this pre-gathering strategy ensures that the memory footprint remains within the bound of that required by model-centric GNN training, as depicted in Figure~\ref{fig:distgnn_default}. For example, when training GAT on the OGB-Products dataset, the model-centric approach requires 530 MB of host memory for temporary feature storage, whereas pre-gathering demands only 87 MB.

%% file: tex/design3.tex
\subsection{Micrograph Merging in GNN Training}
\label{design3}
Micrograph-based GNN training tends to necessitate more frequent GPU kernel
launches and may also entail synchronization overhead at the end of each time
step. Consequently, a trade-off is required between the advantages of
reduced remote feature fetching through model migration and the additional
overhead imposed by micrograph-based training.

\noindent\textbf{Key idea.} By merging micrographs, we can potentially decrease 
the number of time steps 
during the GNN model training, thereby reducing the associated training
overhead. However, the merging process must be approached strategically; random
micrograph consolidation could lead to load imbalance across GPU servers. When
considering the merging of micrographs, two critical questions arise.

\noindent\textbf{Which micrographs to merge?}
Merging micrographs could lead to an increase in remote feature fetching. To
counteract this, we should strategically select micrographs that rely on the
fewest number of remote feature vectors. Additionally, it is imperative to
ensure that different models are concurrently trained on separate servers after
merging. Therefore, all micrographs utilized within a single time step should be
considered for merging. Specifically, for each time step, we calculate the total
count of vertex features, denoted as $Num_{vertex}$, for all micrographs slated
for training. We then pinpoint the time step $ts_{min}$ with the lowest
$Num_{vertex}$ value. However, since we must make decisions before the execution
of an iteration and before micrographs are generated, $Num_{vertex}$ is not yet
determinable. To circumvent this, we approximate $Num_{vertex}$ using the total
number of root vertices, designated for training in a
given time step. Subsequently, we merge the micrographs scheduled for $ts_{min}$
with those from other time steps, ensuring they are used as evenly as possible
by the same model. By doing so, we can balance the time the model takes across
different time steps.
  
\noindent\textbf{How many micrographs should be merged?} If we merge all the
micrographs, micrograph-based training degrades to subgraph-based training. If
we did not merge enough micrographs, the training overhead can be still
significant. Therefore, 
we require an
examination period to determine how many micrographs should be merged. During
this period starting from the second epoch, for each iteration, we identify a
time step $ts_{min}$ and merge the micrographs in the time step with those in
other time steps but for the same model. Then, we measure the execution time of
the current epoch and compare it to that of the previous epoch. If the execution
time is not reduced by merging, we stop the process and use the existing
micrographs for training. Otherwise, we repeat the identifying-and-merging
process until the execution time cannot be reduced. After that, all the
following epochs will use the same merging pattern.

\begin{figure}
        \centering
        \includegraphics[width=3.3in]{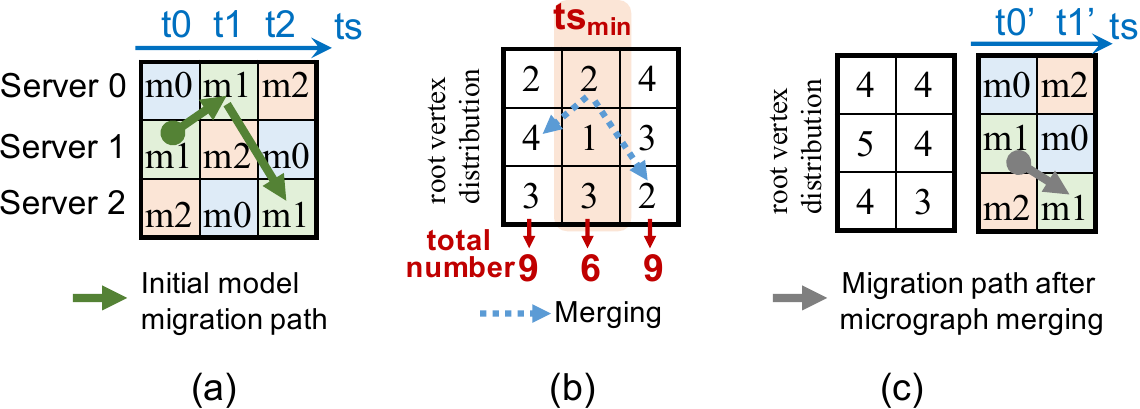}
        \caption{\footnotesize{Illustration of micrograph merging. Squares of
        	the same color indicate the same GNN model. Model migration
        	paths of m2 and m3 are omitted. The total number of root
        	vertices of each model keeps consistent before and after
        	merging.}}
        \label{fig:et_ours}
        \vspace{-0.2in}
\end{figure}

We use an example in Figure~\ref{fig:et_ours} to illustrate the micrograph
merging. We assume that there are three GPU servers including server 0, 1, and
2. Hence, the initial training needs three time steps t0, t1, and t2. We use a
matrix to show the assignment of models (i.e., m0, m1, and m2) across the three
time steps. The initial model distributions and migration paths are shown in
Figure~\ref{fig:et_ours}(a). For merging, we count the total number of
redistributed root vertices for each model at each time step and show them in
Figure~\ref{fig:et_ours}(b). Then, t1 is identified as $ts_{min}$. 
Consequently, we can merge the micrographs from time step t1 with those from
time steps t0 and t2. Taking model m1 for instance, its two root vertices assigned
at time step t1 are evenly distributed across model m1 at time step t0 and t2,
resulting in model m1 having 5 vertices at t0 and 3 at t2. Models m2 and m3
follow a similar redistribution process. After merging, we can remove time step
t1. The revised training process consists of only two time steps t0' and t1'.
The root vertex distribution and model migration paths after merging are shown
in Figure~\ref{fig:et_ours}(c).

%% file: tex/implementation.tex
\section{Implementation}
\label{impl}


We implemented \pname{} based on one of the most popular 
GNN frameworks, DGL~\cite{dgl-arxiv19} (with PyTorch backend). 
We reutilized DGL's graph data 
partitioning module, sampling module, and GNN computation module.
The GNN computation module includes both forward and backward 
propagation, gradient synchronization, and parameter updates.
Our primary focus is on the gathering phase, 
which has been identified as the bottleneck of distributed GNN training.

Before implementing the micrograph-based training, we first developed 
a distributed cache using Golang to store the partitioned graph 
data on each machine. The Python-based GNN application utilizes 
Google Remote Procedure Call (gRPC) to request and retrieve 
vertex features from other machines.
The model migration is implemented using PyTorch's distributed module.
For \techb{}, we utilized a Python list to temporarily store multiple 
micrographs and detected and removed duplicate vertices before 
requesting features from the cache server.
To implement \techc{}, we monitored the 
runtime for each epoch's training and stored its value in a temporary list. 
After that, we utilized this information to adjust
the number of time steps of one iteration as described in~\cref{design3}.

\begin{figure*}
	\centering
	\includegraphics[width=7in, height=1.4in]{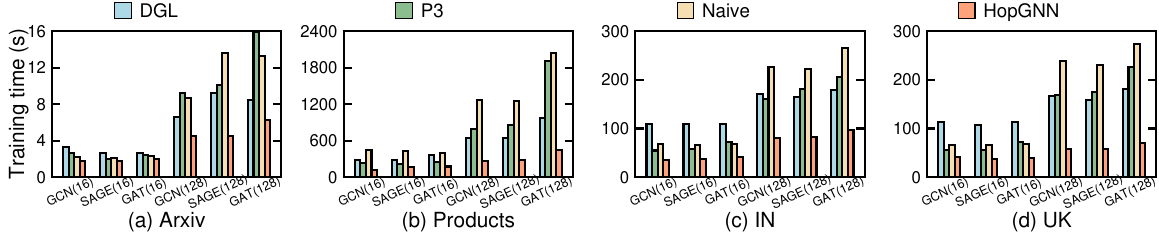}
	\caption{Performance comparison of various training frameworks for three shallow models.}
	\label{fig:overall}
\end{figure*}

\begin{figure}
	\centering
	\includegraphics[width=3.2in, height=1.4in]{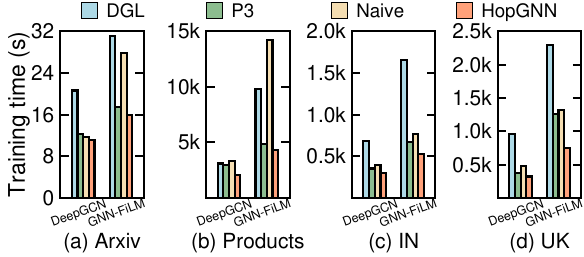}
	\caption{Performance comparison of various training frameworks for two deep models.}
	\label{fig:overall_deepmodel}
	\vspace{-0.2in}
\end{figure}

%% file: tex/eval.tex
\section{Evaluation}
\label{eval}

\input{tex/exp-setup}

\input{tex/exp-overall}

\input{tex/exp-indiv}

\input{tex/exp-reduction}
\input{tex/exp-largedataset}

\input{tex/exp-gpuutil}
\input{tex/exp-fullbatch}
\input{tex/exp-sensi}

\input{tex/exp-acc}

%% file: tex/exp-setup.tex
\subsection{Experimental Setup}
\label{exp:setup}

\begin{table}
	\begin{tabular}{|c|c|c|c|c|c|}
		\hline
		\textbf{Dataset}         & \textbf{\#Vertex} & \textbf{\#Edge} &
		\textbf{Dim.} &\textbf{Vol$_G$} & \textbf{Vol$_F$} \\ \hline
		Arxiv      & 169K     & 1.17M  & 128  &     3.3 MB       &   85 MB
		\\ \hline
		Products   & 2.45M    & 61.9M  & 100  &     464 MB       &     980 MB
		\\ \hline
		UK & 1M     & 41.2M  & 600      &  12 MB    &    2.3 GB        \\ \hline
		IN & 1.38M & 16.9M & 600  &  8.2 MB  & 3.2 GB \\ \hline
		IT & 41.3M & 1.15B & 600 & 363MB & 92.3 GB  \\ \hline
	\end{tabular}
	\caption{The details of graph datasets used in GNN training. \#Vertex and \#Edge denote the number of graph vertices and edges. Dim. denotes the dimension of vertex features. Vol$_G$ and Vol$_F$ denote the data volume sizes of the graph topology and features.}
	\label{tab:datasets}
	\vspace{-0.2in}
\end{table}

\noindent\textbf{System configurations.} We conduct the experiments on a cluster
with four GPU servers, each with 2$\times$Intel(R) Xeon(R) Gold 5318Y CPUs (48
cores), 128 GB CPU memory, and an NVIDIA A100 40GB GPU. All servers are
interconnected with a 10 Gb/s Ethernet network, running Ubuntu v18.04, PyTorch
v1.10.1+cu113, and Python v3.9.0. 

\noindent\textbf{Models and datasets.} We use three shallow models
(G-CN~\cite{gcn-iclr17}, GraphSAGE~\cite{sage-nips17}, GAT~\cite{gat-iclr18})
and two deep models (DeepGCN~\cite{deepgcn-iccv19} and
GNN-FiLM~\cite{film-icml20}) to evaluate \pname{}. Following the
paper~\cite{film-icml20}, we set DeepGCN to include seven layers and GNN-FiLM to
comprise ten layers. Other models all have three layers. Beyond the variation in
the number of layers, these models are distinguished by their distinct methods
for aggregating neighboring vertices. We use 'Model(16)' and 'Model(128)' to
denote the neural network with hidden dimension sizes of 16 and 128
respectively.


We utilize five well-established datasets, detailed in Table~\ref{tab:datasets}.
The Arxiv~\cite{ogb} and Products~\cite{ogb} datasets represent smaller graph
instances, while the UK~\cite{ukindataset} and IN~\cite{ukindataset} datasets
are indicative of medium-scale graphs. In contrast, the IT~\cite{itdataset}
dataset exemplifies a large-scale graph~\cite{itd}. It is important to note that
the original UK, IN, and IT datasets lack vertex features; therefore, we
introduce random features for these datasets, assigning a dimension of 600 to
each vertex, a method akin to those in~\cite{pagraph-socc20, p3-osdi21}. Across
all datasets, we implement a standard neighbor sampling fanout of 10, aligning
with the setup in~\cite{gnnlab-eurosys22}. Owing to the protracted training
durations associated with certain evaluations on the large IT dataset, we limit
our analysis to a select subset of tests, with outcomes presented
in~\cref{sec:large_graph}.

\noindent\textbf{Compared systems.} We evaluate \pname{} against the
industry-leading DGL framework~\cite{dgl-arxiv19} and the state-of-the-art $P^3$~\cite{p3-osdi21} and NeutronStar~\cite{neutronstar-sigmod22} frameworks.
DGL facilitates GNN model training by fetching required features, either locally or remotely. 
$P^3$ integrates model-parallel and data-parallel approaches, minimizing the transfer
of original vertex features but necessitating additional intermediate data
movement. NeutronStar enhances training efficiency by optimizing the balance
between redundant computation and communication time. Unlike these frameworks,
which adopt a "model-centric" approach, \pname{} is "feature-centric".
As $P^3$ is not open-source, we reimplemented it as faithfully as possible
based on the original paper's description. Additionally, we assess a naive
feature-centric approach (Naive) as discussed in \cref{sec:challenges}, to
underscore the value of the techniques introduced in \pname{}. We omit a
direct comparison with ROC~\cite{roc-mlsys20}, given that $P^3$ has previously
surpassed it~\cite{p3-osdi21}. For most experiments, we employ
mini-batch training, excluding NeutronStar due to its limitation to full-batch
training only. However, we present a specific comparison between \pname{} and
NeutronStar in \cref{sec:nstar}.

%% file: tex/exp-overall.tex
\subsection{Overall Performance}
\label{exp:overall}



Figures~\ref{fig:overall} and~\ref{fig:overall_deepmodel} respectively show the end-to-end training times 
for shallow and deep GNN models, due to their significant numerical differences.
We train each model for ten epochs
and report the average training time. We have four observations.

First, \pname{} performs the best among the four frameworks because of the
reduced feature and intermediate data transmissions.
Specifically, it
achieves 1.3--3.1$\times$ speedups over DGL, and 1.2--4.2$\times$ against $P^3$, for the GCN,
GraphSAGE, GAT, DeepGCN, and GNN-FiLM models. 
Compared to Naive, \pname{} obtains up to 4.8$\times$
acceleration. This
demonstrates the effectiveness of
\pname{} across diverse GNN models and datasets.

Second, although sometimes efficient, Naive cannot always deliver performance
improvements over DGL and $P^3$. For example, Naive is
even 1.62$\times$ worse than DGL for GCN(16) on Products. This is because Naive introduces significant intermediate data communication and
frequent model migrations. However, \pname{} always outperforms
the existing frameworks. This shows the naive migration is insufficient and the proposed techniques in \pname{} are necessary.

Third, \pname{}'s speedup varies for different GNN models. It
achieves 2.5$\times$ acceleration for GCN whereas 2.2$\times$ for GAT on Products. 
This variation mainly
arises from the varying time proportion of feature gathering in the training, 
resulting in different potential for performance improvements. 
For example, as GAT involves
attention-based aggregation whereas GCN
employs a simpler summation aggregation, the remote feature gathering times of
these two models account for 50.3\% and 39.1\% of
their training times, respectively.  

Fourth, \pname{}'s improvement is independent of the hidden dimension of
the models, while $P^3$'s speedup is sensitive to this~\cite{p3-osdi21, bgl-nsdi23}.
For example, with GAT on the IN dataset, when the hidden dimension size is 128,
$P^3$ is 1.2$\times$ slower than DGL, while \pname{} still achieves 
1.8$\times$ performance improvement.
This is because \pname{}
performs the forward and backward propagation of a micrograph on a single
server, eliminating the inter-server transmission of hidden embeddings of $P^3$.

%% file: tex/exp-indiv.tex
\subsection{Impact of Individual Techniques}
\label{exp:indiv}


\begin{figure}
	\centering
	\includegraphics[width=3.3in, height=1.4in]{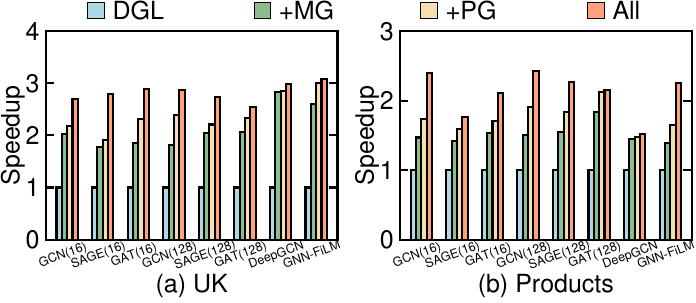}
	\caption{Improvements of each individual technique. The training time of DGL is normalized to one.}
	\label{fig:indiv_all}
	\vspace{-0.2in}
\end{figure}

Figure~\ref{fig:indiv_all} shows the impact of each technique on the
per-epoch training performance. We enable each technique based on DGL. \textit{+MG} denotes the
version where \techa{} is turned on. \textit{+PG} denotes the version where
\techb{} is added based on \textit{+MG}. \textit{All} means the \techc{} is also
enabled. Due to the space
limitation and similar trends on other datasets, we only present the results on Products and UK. We have
three observations.

First, the speedup increases with the incorporation of each
technique of \pname{}. For example, for GAT,
\textit{+MG}, \textit{+PG}, and \textit{All}
achieve improvements of up to 1.69$\times$, 1.92$\times$, and 2.14$\times$
against DGL on Products, and 1.96$\times$,
2.33$\times$, and 2.72$\times$ on UK. This is because the \techa{}
can reduce the number of remote feature gathering, \techb{} can further reduce
the transmission of redundant features, and \techc{} can adaptively
decrease the model migration frequency based on model and dataset
characteristics.

\begin{figure}
	\begin{adjustbox}{valign=c}
		\begin{minipage}{0.42\linewidth}
			\flushleft
			\scalebox{0.8}{
				\begin{tabular}{|c|c|c|}
					\hline
					&         & Miss Rate \\ \hline
					\multirow{2}{*}{Arxiv}    & DGL  &  74\%   \\ \cline{2-3} 
					& +MG      &   43\%   \\ \hline
					\multirow{2}{*}{Products} & DGL & 77\%   \\ \cline{2-3} 
					& +MG      & 22\%   \\ \hline
					\multirow{2}{*}{UK} & DGL &    78\%       \\ \cline{2-3} 
					& +MG      &    19\%       \\ \hline
					\multirow{2}{*}{IN} & DGL &     77\%      \\ \cline{2-3} 
					& +MG      &    9.2\%       \\ \hline
				\end{tabular}
			}
			\caption{Miss rates.}
			\label{tab:indiv_techa}
		\end{minipage}
	\end{adjustbox}
	\hfill
	\begin{adjustbox}{valign=c}
		\begin{minipage}{0.5\linewidth}
			\flushright
			\includegraphics[width=1.6in,height=1.2in]{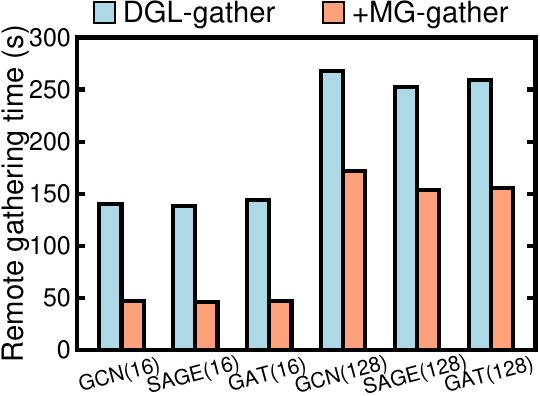}
			\caption{Details after using micrograph-based training.}
			\label{fig:indiv_techa}
		\end{minipage}
	\end{adjustbox}%
\end{figure}

Second, the first technique provides the most pronounced performance
improvement: it achieves 74\% improvement on average while the second and third obtain 11\% and
15\%. This is because it significantly
decreases the average miss rate of feature gathering from 76.5\% to 23.3\%
across the four datasets (see Figure~\ref{tab:indiv_techa}), which can
effectively address the data transmission bottleneck in most cases.

Third, each technique brings different benefits across
data-sets. For example, the first technique achieves an average improvement of 1.52$\times$ on Products whereas 2.13$\times$ on UK.
This variation arises from the differences in vertex feature dimensions and
topologies in different datasets.

\begin{figure}
	\begin{adjustbox}{valign=c}
		\begin{minipage}{0.45\linewidth}
			\flushleft
			\includegraphics[width=1.5in, height=1in]{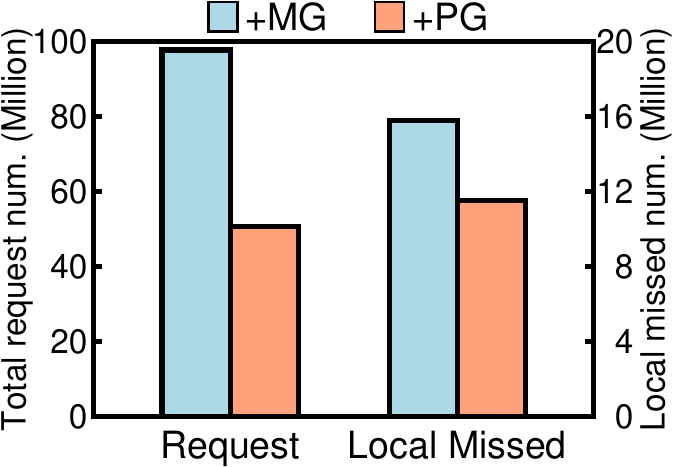}
			\caption{Details after using \techb{}.}
			\label{fig:indiv_techb}
		\end{minipage}
	\end{adjustbox}
	\hfill
	\begin{adjustbox}{valign=c}
		\begin{minipage}{0.45\linewidth}
			\centering
			\includegraphics[width=1.6in, height=1in]{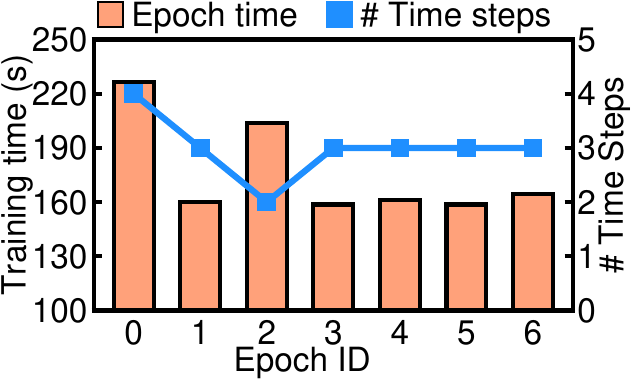}
			\caption{Details after using \techc{}.}
			\label{fig:indiv_techc}
		\end{minipage}
	\end{adjustbox}%
	\vspace{-0.1in}
\end{figure}




Figure~\ref{fig:indiv_techa} presents the remote feature gathering time for the
Products dataset with and without enabling the \techa{}. It shows that our
technique significantly reduces the gathering time 2.3$\times$ on average, thus 
reducing the overall training time.

Figure~\ref{fig:indiv_techb} shows the number of remote feature requests and the
number of local feature miss requests when enabling the \techb{}.
The result indicates that the \techb{} further reduces the former by 1.9$\times$
and the latter 1.4$\times$, showing the efficiency of the \techb{}.


Figure~\ref{fig:indiv_techc} illustrates the epoch time and the corresponding
number of time steps per iteration for the GAT model on the Products dataset
when employing the \techc{} technique. Initially, with four available machines,
the epoch 0 begins with four time steps. Subsequently, \pname{} dynamically
optimizes the number of time steps, reducing it to three in epoch 1 and further
to two by epoch 2. Ultimately, \pname{} determines that utilizing three time
steps per iteration for the remainder of the training, resulting in the most
efficient training time.

\begin{figure}
	\begin{minipage}[b]{0.45\linewidth}
		\centering
		\subfigure[Training time.]{
			\includegraphics[width=1.5in, height=1.2in]{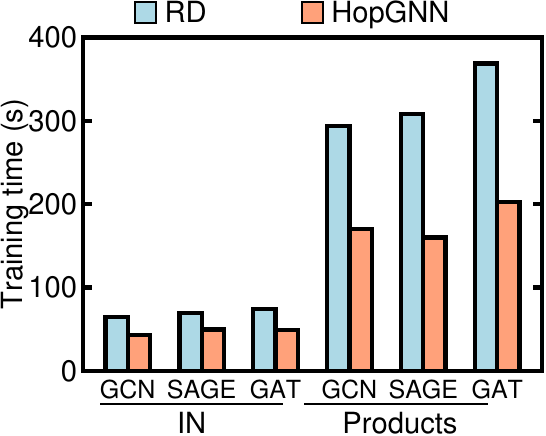}
			\label{fig:indiv_ms:a}
		}
	\end{minipage}
	\hfill
	\begin{minipage}[b]{0.45\linewidth}
		\subfigure[Workload distribution.]{
			\includegraphics[width=1.5in, height=1.2in]{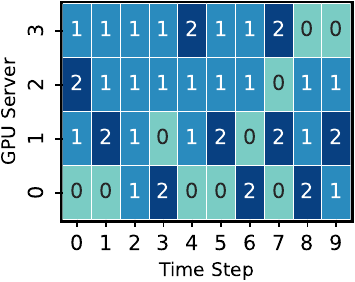}
			\label{fig:indiv_ms:b}
		}
	\end{minipage}
	\vspace{-0.1in}
	\caption{The impact of selection schemes.}
	\vspace{-0.1in}
\end{figure}

\begin{figure}
	\begin{adjustbox}{valign=c}
		\begin{minipage}{0.45\linewidth}
			\flushleft
			\includegraphics[width=1.6in, height=1.1in]{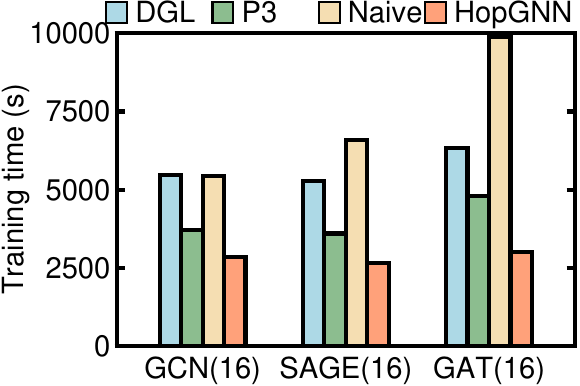}
			\caption{Performance on the IT graph dataset.}
			\label{fig:large_graph}
		\end{minipage}
	\end{adjustbox}
	\hfill
	\begin{adjustbox}{valign=c}
		\begin{minipage}{0.45\linewidth}
			\flushright
			\includegraphics[width=1.5in, height=1.2in]{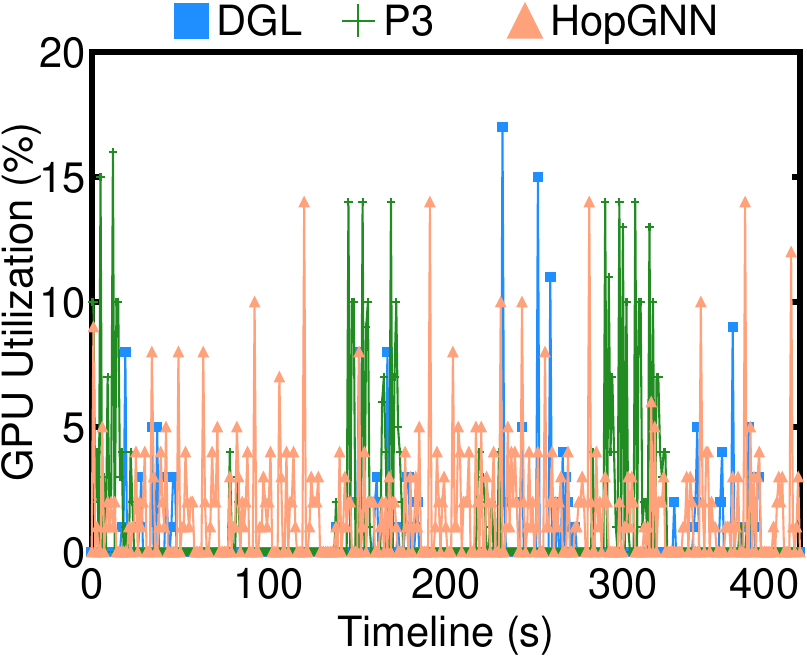}
			\caption{GPU utilization with different systems.}
			\label{fig:gpu_util}
		\end{minipage}
	\end{adjustbox}%
	\vspace{-0.2in}
\end{figure}

%% file: tex/exp-reduction.tex
\subsection{Micrograph Merging Selection}
To demonstrate the effectiveness of our selection method (\cref{design3}) in the
\techc,
we compare it with a random scheme (RD), where a micrograph is randomly selected
to merge with other micrographs for each model.
Figure~\ref{fig:indiv_ms:a} illustrates that our method outperforms RD by
1.4--1.9$\times$ on IN and Products. 
Figure~\ref{fig:indiv_ms:b} shows the number of GNN training models on each GPU
server at each time step with RD. It shows that RD has an
uneven workload distribution among servers, thereby degrading the training performance.


%% file: tex/exp-largedataset.tex
\subsection{Results on Large-Scale Graph}
\label{sec:large_graph}

Figure~\ref{fig:large_graph} illustrates the epoch training time of different
systems on the large dataset IT.
Due to the extensive training times, we only conduct a subset of the
tests. 
\pname{} achieves an average acceleration of 1.91$\times$ and 1.48$\times$ against DGL and $P^3$, respectively. 
The improvement is attributed to the increase in the local feature 
hit rate from 24.4\% to 92.3\% after employing the techniques in \pname{}.
This result shows \pname{} is still effective on the large dataset.

%% file: tex/exp-gpuutil.tex
\subsection{GPU Utilization}
Figure~\ref{fig:gpu_util} illustrates the GPU utilization of \pname{}, DGL, and
$P^3$ for the GAT model on the UK dataset. 
Similar results are observed on other models and datasets. 
We utilize the Python library GPUtil~\cite{gputil} (which relies on nvidia-smi) to capture the GPU utilizations every 250ms during a steady 400-second running time window. 
We observe that the peak GPU utilization is smaller than 20\% in all these systems due to the sparse nature of computations~\cite{p3-osdi21} and the high speed of A100.
However, \pname{} is able to keep GPU busy (i.e., at least one core active) for 52\% of the total time, while DGL and $P^3$ only achieve 13\% and 18\%, respectively.
This explains why \pname{} achieves the shortest training time.

%% file: tex/exp-fullbatch.tex
\subsection{Comparison with NeutronStar}
\label{sec:nstar}
\begin{figure}
	\centering
	\subfigure[IN dataset.]{
		\includegraphics[width=1.5in, height=1.1in]{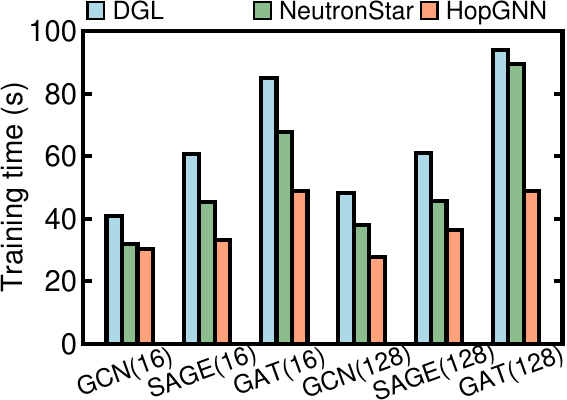}
	}
	\hfill
	\subfigure[UK dataset.]{
		\includegraphics[width=1.5in, height=1.1in]{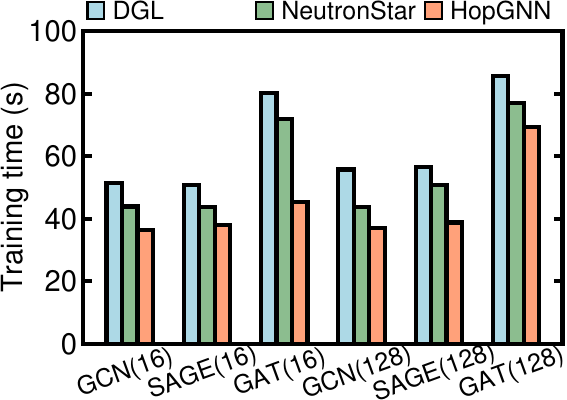}
	}
	\caption{Performance comparison with full-batch training.}
	\label{fig:compare_nstar}
	\vspace{-0.1in}
\end{figure}


Since NeutronStar does not support sampling, we disable sampling in all compared
systems. For fair comparison, we reproduce NeutronStar based on the DGL
framework. As Figure~\ref{fig:compare_nstar} shows, both NeutronStar and
\pname{} are more efficient than DGL, with \pname{} performing the best.
\pname{} achieves a speedup of 1.05--1.82$\times$ over NeutronStar. This is
because although NeutronStar accelerates DGL by reducing redundant computations,
the proportion of computation is smaller than that of feature communication in
our test scenario. Therefore, \pname{} has a shorter overall training time by
reducing feature communication through model migration.

%% file: tex/exp-sensi.tex
\subsection{Sensitivity Analysis}
\label{exp:sensi}


\begin{figure}
	\centering
	\subfigure[Various batch sizes.]{
		\includegraphics[width=1.5in, height=1.0in]{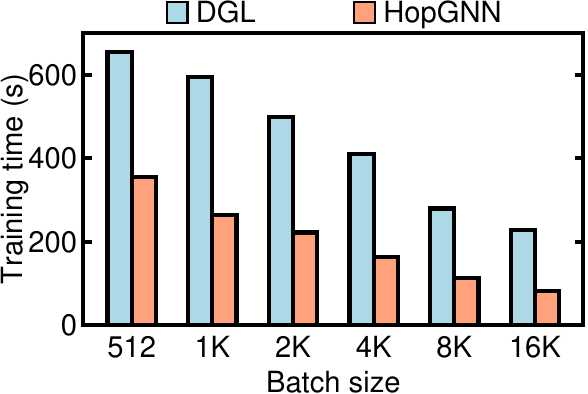}
		\label{fig:sens_bs}
	}
	\hfill
	\subfigure[Various dimensions of features.]{
		\includegraphics[width=1.5in, height=1.0in]{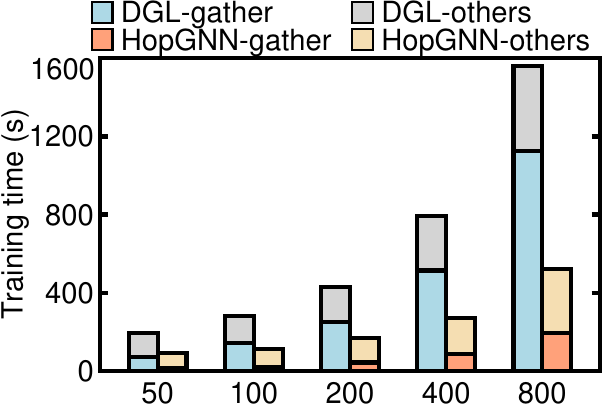}
		\label{fig:sens_featdim}
	} 
	\caption{Performance comparison with various batch sizes and feature dimensions.}
	\vspace{-0.2in}
\end{figure}



\noindent\textbf{Batch size.}
Figure~\ref{fig:sens_bs} depicts the training time of GCN on Products with various
batch sizes. \pname{} consistently outperforms DGL for
batch sizes from 512 to 16K, with performance improvements of 2.2--2.8$\times$.
This can be mainly attributed to \pname{}'s ability to reduce remote feature
fetching time.

\noindent\textbf{Feature dimension.}
Figure~\ref{fig:sens_featdim} shows the system performance on Products with
different feature dimensions. As the feature dimension increases, \pname{}'s
speedup is increased from 2.1$\times$ to 2.9$\times$. This is because the
proportion of remote feature gathering time in DGL rises from 36.8\% to
72.0\%, leaving more space for acceleration with \pname{}.

\begin{figure}
	\begin{minipage}[b]{0.45\linewidth}
		\centering
		\subfigure[Various fanouts.]{
			\includegraphics[width=1.5in, height=1.0in]{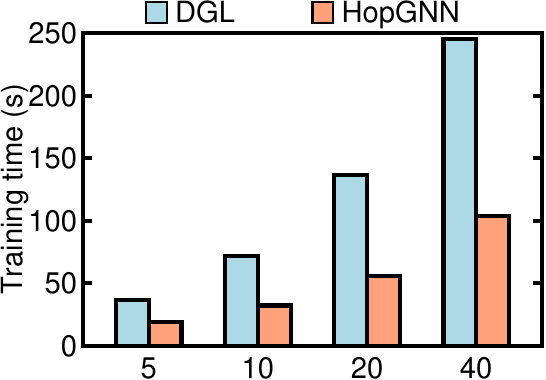}
			\label{fig:sens_fanout}
		} 
	\end{minipage}
	\hfill
	\begin{minipage}[b]{0.45\linewidth}
		\subfigure[Various number of machines.]{
			\includegraphics[width=1.5in, height=1.0in]{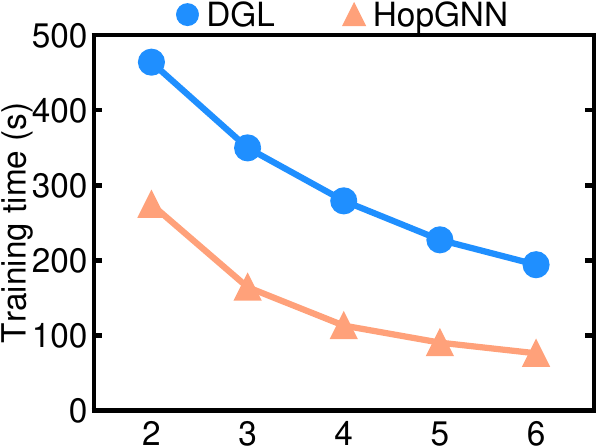}
			\label{fig:sens_nmachine}
		}
	\end{minipage}
	\caption{Performance comparison with various fanouts and number of machines.}
	\vspace{-0.1in}
\end{figure}

\noindent\textbf{Fanout size.}
Figure~\ref{fig:sens_fanout} shows the system performance of \pname{} with
different fanouts. \pname{} consistently outperforms DGL by
2.3$\times$ on average. Furthermore, it offers better scalability than DGL
on high-dimensional large graph datasets. Specifically, when the feature
dimension expands by a factor of 8 (from 5 to 40), \pname{}'s training time is
increased by 5.3$\times$ while DGL's time is increased by 6.6$\times$.


\noindent\textbf{Number of distributed machines.}
Figure~\ref{fig:sens_nmachine} presents the system performance of GCN on Products with various number of machines. We observe that \pname{} consistently
outperforms DGL by 2.27$\times$ on average. Furthermore, as the number of
machines increases from 2 to 6, \pname{}'s speedup is increased from 1.69$\times$ to
2.55$\times$. This shows \pname{} has better scalability than
DGL in multi-machine scenarios.

\begin{table}
	\begin{tabular}{|c|c|c|cc|cc|}
	 \hline
	 \multirow{2}{*}{\textbf{\begin{tabular}[c]{@{}c@{}}Dat-\\aset \end{tabular}}} & \multirow{2}{*}{\textbf{Models}} & \multirow{2}{*}{\textbf{\begin{tabular}[c]{@{}c@{}}DGL\\ Acc.\end{tabular}}} & \multicolumn{2}{c|}{\textbf{LO}}                   & \multicolumn{2}{c|}{\textbf{\pname{}}}                 \\ \cline{4-7} 
	 &                                  &                                                                                  & \multicolumn{1}{c|}{\textbf{Acc.}} & \textbf{Drop} & \multicolumn{1}{c|}{\textbf{Acc.}} & \textbf{Drop} \\ \hline
	 \multirow{3}{*}{Arxiv}            & GCN                              & 60.24                                                                            & \multicolumn{1}{c|}{59.71}         & \textbf{0.53} & \multicolumn{1}{c|}{60.24}         & \textbf{S}    \\ \cline{2-7} 
	 & SAGE                        & 61.96                                                                            & \multicolumn{1}{c|}{61.96}         & \textbf{S}    & \multicolumn{1}{c|}{61.98}         & \textbf{S}    \\ \cline{2-7} 
	 & GAT                              & 59.28                                                                            & \multicolumn{1}{c|}{59.15}         & \textbf{0.13} & \multicolumn{1}{c|}{59.25}         & \textbf{S}    \\ \hline
 	\end{tabular}
	\caption{Model accuracy (\%). The column in bold indicates
	 accuracy drop. "S" stands for "same", indicating
	 that the accuracy drop is within 0.1\%.}
	\label{tab:accdrop}
	\vspace{-0.3in}
   \end{table}


%% file: tex/exp-acc.tex
\subsection{Model Accuracy}
\label{exp:acc}

So far we have compared \pname{} with the state-of-the-art systems with accuracy fidelity.
Approximate methods, such as selectively ignoring remote vertex
features~\cite{distgnn-sc21}, proximity-aware ordering~\cite{bgl-nsdi23}, and
locality-optimized method (LO) \cite{pagraph-socc20, distdgl-ia320}, have also been proposed to
reduce remote feature gathering time. 
However, these systems may compromise 
model accuracy.
For example, \cite{distgnn-sc21} has demonstrated a 0.95\%
accuracy drop for SAGE on Products and \cite{bgl-nsdi23} has shown a 0.2\%
accuracy drop on the OGB-Papers dataset~\cite{ogb}.
Since the impact of LO on GNN accuracy has not been studied, we conduct tests to study this on the Arxiv
dataset. 
Table~\ref{tab:accdrop} shows \pname{} maintains the same accuracy as DGL 
while LO does not. 
This is because LO only chooses the vertices from the local node, 
introducing bias into the
training sequence, potentially degrading the model's accuracy, as
discussed in \cite{globally-ipdps22,accele-hipc19}.
Given that 0.1\% accuracy loss could lead to substantial economic consequences~\cite{persia-kdd22, under-hpca21,
checknrun-ndsi22,aibox-cikm19}
 and our work is able to maintain
accuracy fidelity, 
we do not compare \pname{}'s training time with systems
that may compromise accuracy~\cite{distgnn-sc21, bgl-nsdi23,
distdgl-ia320, pagraph-socc20}.

%% file: tex/discussion.tex
\section{Discussion}
\label{disscus}

\noindent\textbf{Graph partitioning time.}  
While the METIS graph partitioning algorithm used by \pname{} is more time-consuming than
$P^3$'s random partitioning method,
it runs offline and only once. Thus, its partitioning time can be amortized 
over the large number of training epochs and GNN tasks. 
For example, when partitioning a large IT graph, although \pname{}  
takes approximately 2800 seconds longer than random partitioning,
it still outperforms $P^3$ by 1.6$\times$ on GAT in a typical 200-epoch training
scenario, with partitioning time included.
Moreover, several recent GNN-tailored partitioning algorithms~\cite{bgl-nsdi23, bytegnn-vldb22, distgnn-sc21, pagraph-socc20} 
have shown potential to further reduce partitioning durations 
while maintaining high locality.

\noindent\textbf{Time and space overhead.} 
\pname{} incurs additional communication time due to the migration of models and gradients across servers compared to DGL. However, the added time
overhead is negligible, averaging only 4.6\% of the total training time. As shown
in \cref{exp:overall}, its training time (including the communication) still outperforms other approaches. Regarding memory usage, despite
receiving multiple partial gradients per model during each iteration,
\pname{}  maintains memory efficiency equivalent to DGL. 
This is because \pname{} accumulates incoming partial gradients with the 
existing ones and updates them in place.

\noindent\textbf{Failure recovery.} 
In \pname{}, the GNN model may reside on various servers at different time
steps. To facilitate resumption from the last completed time step after a server
failure, a straightforward approach is to checkpoint iteration ID, time step ID, model IDs, accumulated partial gradients, and model
parameters at every time step. However, to enhance checkpointing efficiency, we
opt for an iteration-level checkpointing strategy. This method necessitates that
each model only keeps track of its iteration ID and model parameters at selected
intervals, as the accumulated partial gradients are cleared at the end of
each iteration. Moreover, this technique aligns with established
checkpointing strategies and can be seamlessly integrated with advanced
optimizations, such as asynchronous checkpointing~\cite{checkfreq-fast21},
potentially enhancing overall system efficiency.

\noindent\textbf{Generality of \pname{}.}
Since \pname{} does not modify the GNN computation kernel functions 
and performs forward and backward computations for each micrograph on a single server, 
it maintains robust compatibility with GNN models that use different aggregation 
operations (e.g., sum, max, LSTM) as supported by the original DGL framework.
The efficacy of \pname{} stems from the locality of the micrograph, 
rendering it unsuitable for random graph partitioning algorithms~\cite{p3-osdi21}. 
However, most of GNN partitioning algorithms provide strong
locality \cite{metis98, dgcl-eurosys21, neugraph-atc19,pagraph-socc20,
bgl-nsdi23, bytegnn-vldb22}, endowing \pname{} with excellent practicality and
applicability.


\vspace{-0.15in}


%% file: tex/related.tex
\section{Related Work}
\label{related}
\noindent\textbf{Graph partitioning optimizations.}  DGL~\cite{dgl-arxiv19} utilizes the
METIS graph partitioning algorithm to minimize the number of cut edges.
ByteGNN~\cite{bytegnn-vldb22} and BGL~\cite{bgl-nsdi23} considers multiple-hop
neighbors to further reduce cross-machine vertex accesses.
ROC~\cite{roc-mlsys20} proposes an online linear regression model to optimize
graph partitioning. These efforts aim to maximize the co-location of adjacent
vertices on the same machine, thereby reducing inter-machine feature transmission. 
They are orthogonal to our work and could potentially enhance \pname{}'s performance.

\noindent\textbf{Sampling algorithm optimizations.}
These works focus on improving the locality of feature accesses by changing
sampling algorithms~\cite{pagraph-socc20, distdgl-ia320, bgl-nsdi23,
distgnn-sc21}. As mentioned before, these works tend to compromise the
randomness of GNN sampling, thereby impacting GNN training accuracy. In
contrast, \pname{} has the model accuracy fidelity.

\noindent\textbf{Cache optimizations.} These studies aim to design GPU memory caches to
reduce the feature fetching time from CPU memory. 
PaGraph~\cite{pagraph-socc20} and GNNLab~\cite{gnnlab-eurosys22} implement
static caches to store features of vertices with the highest degree or access
frequency. BGL~\cite{bgl-nsdi23}, on the other hand, employs a dynamic FIFO
cache to balance cache management overhead with hit rate efficiency.
Legion~\cite{legion-atc23} contributes a unified multi-GPU cache strategy to
reduce topology and feature transmissions over PCIe. These methods leverage
additional GPU memory and are complementary to our research. We posit that
integrating these techniques into \pname{} could significantly enhance training
performance.

\noindent\textbf{Computation optimizations.}
Considerable research efforts \cite{mgg-osdi23, gnnadvisor-osdi21, understanding-ppopp21,
neugraph-atc19} have been made to accelerate GNN computation via fine-grained
pipeline and balanced workload scheduling across multiple GPU cores. They are
designed for small graphs that can fit entirely into GPU memory. In contrast,
\pname{} focuses on large graphs with distributed GNN training.

\noindent\textbf{Others system optimizations.} 
DGCL~\cite{dgcl-eurosys21} aims to enhance GPU-to-GPU communication efficiency by employing a multi-path selection algorithm at the physical communication link level. Betty~\cite{betty-asplos23} addresses the challenge of training large batches of data on a single GPU through batch splitting techniques. Dorylus~\cite{dorylus-osdi21} optimizes communication for serverless training scenarios with Lambda servers, a configuration not present in our system. These approaches are orthogonal to and can complement \pname{}.
\vspace{-0.15in}

%% file: tex/conclusion.tex
\section{Conclusion}
\label{conclusion}

In this paper, we present \pname{}, a feature-centric distributed
GNN training framework to reduce inter-machine communication overhead.
\pname{} moves the GNN model
towards the training features, rather than moving the features to the models as in
the existing model-centric frameworks.
To tackle the challenges of implementing this approach, we propose the \techa{},
vertex feature \techb{}, and \techc{}, to reduce remote feature
fetching, intermediate data, and synchronization overhead over network.
Our experimental results demonstrate that \pname{} can achieve a speedup of up to 4.2$\times$ compared to the state-of-the-art framework, $P^3$, across a variety of GNN models and datasets.